\documentclass[prb,showpacs,preprint]{revtex4}
\usepackage{amsmath}
\usepackage{graphicx}
\usepackage{epsfig}

\begin{document}
\title{Antiferromagnetic coupling of the single-molecule magnet Mn$_{12}$
to a ferromagnetic substrate}
\author{Kyungwha Park}
\affiliation{Department of Physics, Virginia Tech, Blacksburg, Virginia, 24061}
\begin{abstract}
We investigate magnetic coupling between a monolayer of prototype single-molecule magnets Mn$_{12}$ 
and a ferromagnetic Ni(111) substrate through S, using density-functional theory (DFT) and a DFT+$U$ 
method. Our DFT and DFT+$U$ calculations show that the Mn$_{12}$ molecules favor antiferromagnetic 
coupling to the Ni substrate, and that they possess magnetic moments deviated
from the magnetic moments of isolated Mn$_{12}$ molecules. We find that the magnetic easy axis 
of the Mn$_{12}$ on Ni (whole system) is dictated by that of the Ni substrate. 
The antiferromagnetic coupling is, dominantly, 
caused by superexchange interactions between the magnetic moments of the Mn and the Ni substrate 
via the S, C, and O anions. 
Our findings can be observed from x-ray magnetic circular dichroism or scanning tunneling
microscopy.
\end{abstract}
\date{\today}

\maketitle

\section{Introduction}

Recently, tailoring static and dynamic magnetic properties of individual atoms and molecules by local 
environmental factors or local probes has drawn a lot of attention. This aligns with an effort to 
develop efficient devices for magnetic storage, spin transfer, molecular spintronics, or quantum 
computation, based on nanometer-sized magnetic atoms or molecules. One promising candidate for such 
devices is a group of single-molecule magnets (SMMs) consisting of several transition metal ions interacting 
through ligands. An individual SMM has a large magnetic moment and a high magnetic anisotropy barrier 
caused by spin-orbit coupling (SOC) and Jahn-Teller distortions \cite{FRIE96,CHUD98}. In order to build 
devices including magnetic molecules, stable monolayers of magnetic molecules must be formed on various 
substrates, and mechanisms of interactions between molecules and substrates should be understood. 
A great progress has been made for deposition of various SMMs, such as prototype SMM Mn$_{12}$ 
\cite{CORN03,FLEU05,SALM07,VOSS07,VOSS08,MANN09}, Mn$_6$ \cite{Mn6}, Fe$_4$ \cite{Fe4}, Cr$_7$Ni \cite{Cr7Ni}, 
and TbPc$_2$ \cite{TbPc2}, on metallic, semiconducting, or superconducting substrates \cite{CORO10}. 
Electronic and magnetic properties of SMMs on substrates were characterized 
using x-ray absorption and photoemission spectroscopy \cite{FLEU05,VOSS07,Mn6,Fe4,Cr7Ni}, 
scanning tunneling microscopy (STM) \cite{CORN03,VOSS08,TbPc2}, $\beta$-detected nuclear magnetic 
resonance \cite{SALM07}, and x-ray magnetic circular dichroism (XMCD) \cite{MANN09,Fe4}. On the 
theoretical front, the electronic structure and magnetic properties of a SMM Mn$_{12}$ monolayer 
adsorbed on Au were studied using density-functional theory (DFT) \cite{SALV07}. 
The theoretical calculations \cite{SALV07} revealed that the Mn$_{12}$ molecules are weakly coupled 
to the Au substrate, and that charge and spin transfer occurs from the Au substrate to the Mn$_{12}$, 
mainly, through linker molecules.

Despite the great advances made in recent years, studies of SMMs adsorbed on substrates are currently 
limited to non-magnetic substrates. An understanding of magnetic coupling between SMMs and ferromagnetic (FM) 
substrates is, however, of a great importance for device applications and for investigation of properties 
of SMMs. For instance, the magnetic coupling can be used to control the reversals of the magnetic moments of 
SMMs or of the magnetization of FM substrates, depending on the coercive field strength of the SMMs 
relative to that of the FM substrates, similarly to exchange-bias systems \cite{NOGU05}. Additionally, 
the magnetic coupling can influence electron transport properties through SMMs bridged between FM electrodes.

In this work, we examine magnetic coupling between a monolayer of SMMs Mn$_{12}$ and a FM Ni(111) substrate, 
using DFT and a DFT+U method \cite{ANIS97,DUDA98}. We consider a structure where SMMs Mn$_{12}$ are adsorbed 
on a Ni substrate via two S atoms. Our DFT and DFT+U calculations show that the Mn$_{12}$ molecules prefer 
antiferromagnetic (AFM) coupling to the Ni substrate, and that their magnetic moments differ
from the magnetic moments of isolated Mn$_{12}$ molecules. We find that the magnetic easy axis of the 
adsorbed Mn$_{12}$ is determined by the magnetic easy axis of the Ni substrate which depends on a thickness 
of the substrate. We also clarify mechanisms of the magnetic coupling by taking into account direct exchange 
coupling such as hybridization and an indirect exchange coupling mechanism such as superexchange interactions
\cite{GOOD63}. We find that superexchange interactions prevail over direct exchange in the magnetic coupling. 
Our results can be observed in experiments such as XMCD or STM.

The outline of this paper is as follows. We introduce our computational methods in Sec.II. We systematically 
construct a Mn$_{12}$ monolayer adsorbed on a Ni substrate via S, as well as present properties of an 
isolated Mn$_{12}$ and a bare Ni slab in Sec.III. We discuss the sign, magnitude, and mechanisms of the 
magnetic coupling between the Mn$_{12}$ and the Ni substrate, and the electronic and magnetic properties 
of the system of interest in Sec.IV. Finally, we make our conclusion in Sec.V.

\section{Computational Methods}

We use two DFT codes, {\tt SIESTA} \cite{SIESTA} and {\tt VASP} \cite{VASP}, to compute the electronic structure 
and magnetic properties of a Mn$_{12}$ monolayer on Ni(111). {\tt SIESTA} uses localized numerical atomic orbitals 
as basis sets, while {\tt VASP} uses plane waves. In both {\tt SIESTA} and {\tt VASP} calculations, we use 
the Perdew-Burke-Ernzerhopf (PBE) generalized-gradient approximation (GGA) \cite{PERD96} for exchange-correlation
potential. We do not include SOC in our calculations unless specified otherwise. First, we discuss set-ups and 
parameters for {\tt SIESTA} calculations. We use Troullier-Martins pseudopotentials \cite{TROU91,ABINIT} with 
scalar relativistic terms and core corrections for all the elements except H. We construct corresponding 
basis sets as discussed in Refs.[\onlinecite{JUNQ01,SALV09}]. For Mn, we include 3$p$ orbitals in valence 
states. For Ni, we use a default basis set of DZP. We use a mesh cutoff of 400 Ry. We carry out 
self-consistent calculations until a density matrix converges to within $3 \times 10^{-5}$. Second, we discuss set-ups 
and parameters for {\tt VASP} calculations. We use projector-augmented-wave (PAW) pseudopotentials \cite{PAW} for all 
the elements. For Mn, we consider 3$p$ orbitals as a semicore. For Ni, valance states consist of 4$s$ and 3$d$ orbitals. 
We use a kinetic energy cutoff of 400~eV. We perform self-consistent calculations until the total 
energy converges to within $1 \times 10^{-4}$~eV. For DFT+U calculations \cite{ANIS97}, we use an on-site Coulomb 
repulsion $U$ term of 4~eV \cite{SALV08} for Mn $d$ orbitals only. The value of $U$ is selected in order to reproduce 
photoemission spectra measured for a crystal of Mn$_{12}$ molecules \cite{PENN06}.

\section{Whole structure}

\begin{figure}
\includegraphics[width=9.cm, height=8.cm]{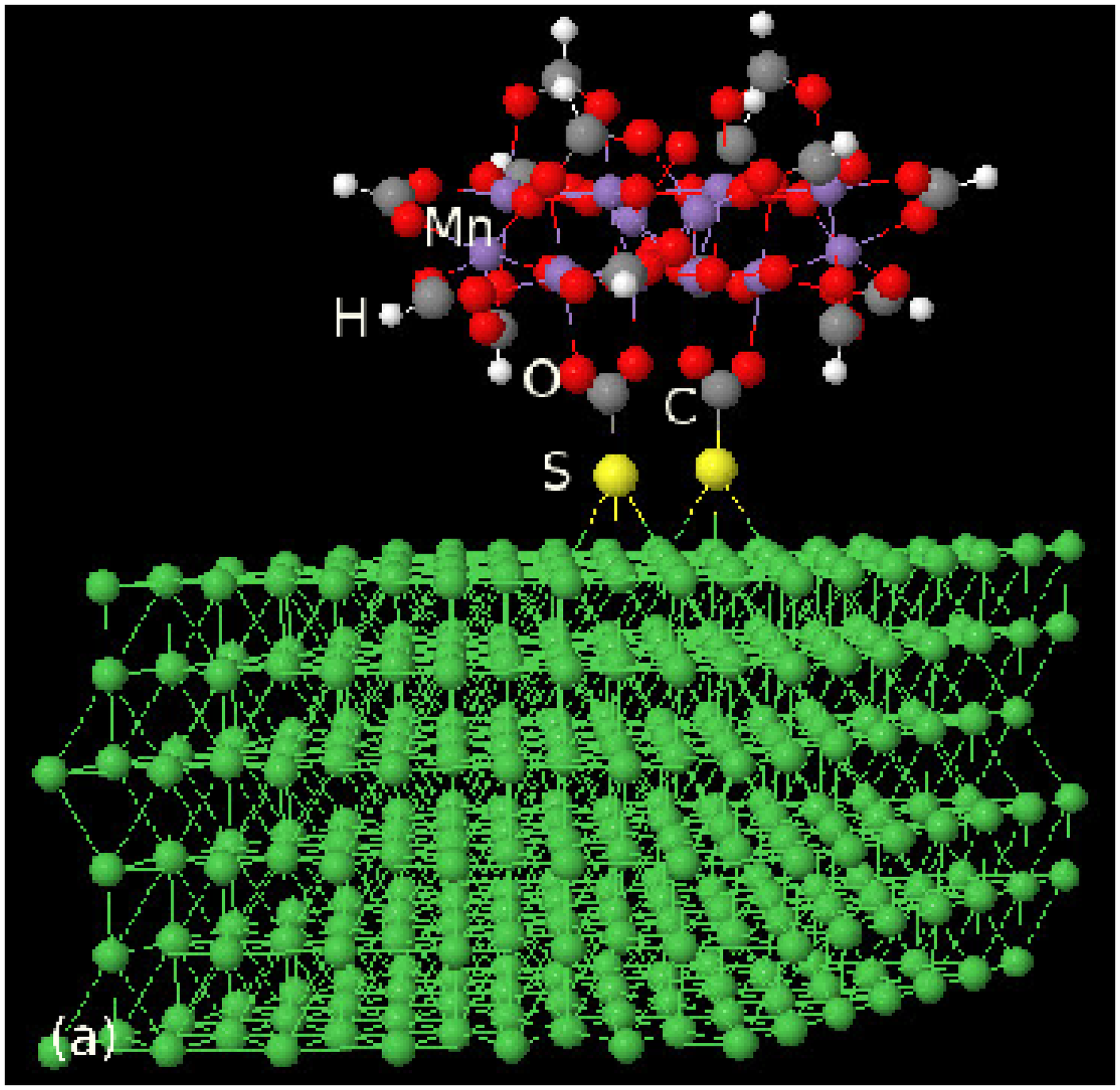}
\hspace{1.5truecm}
\includegraphics[width=4.cm, height=8.cm]{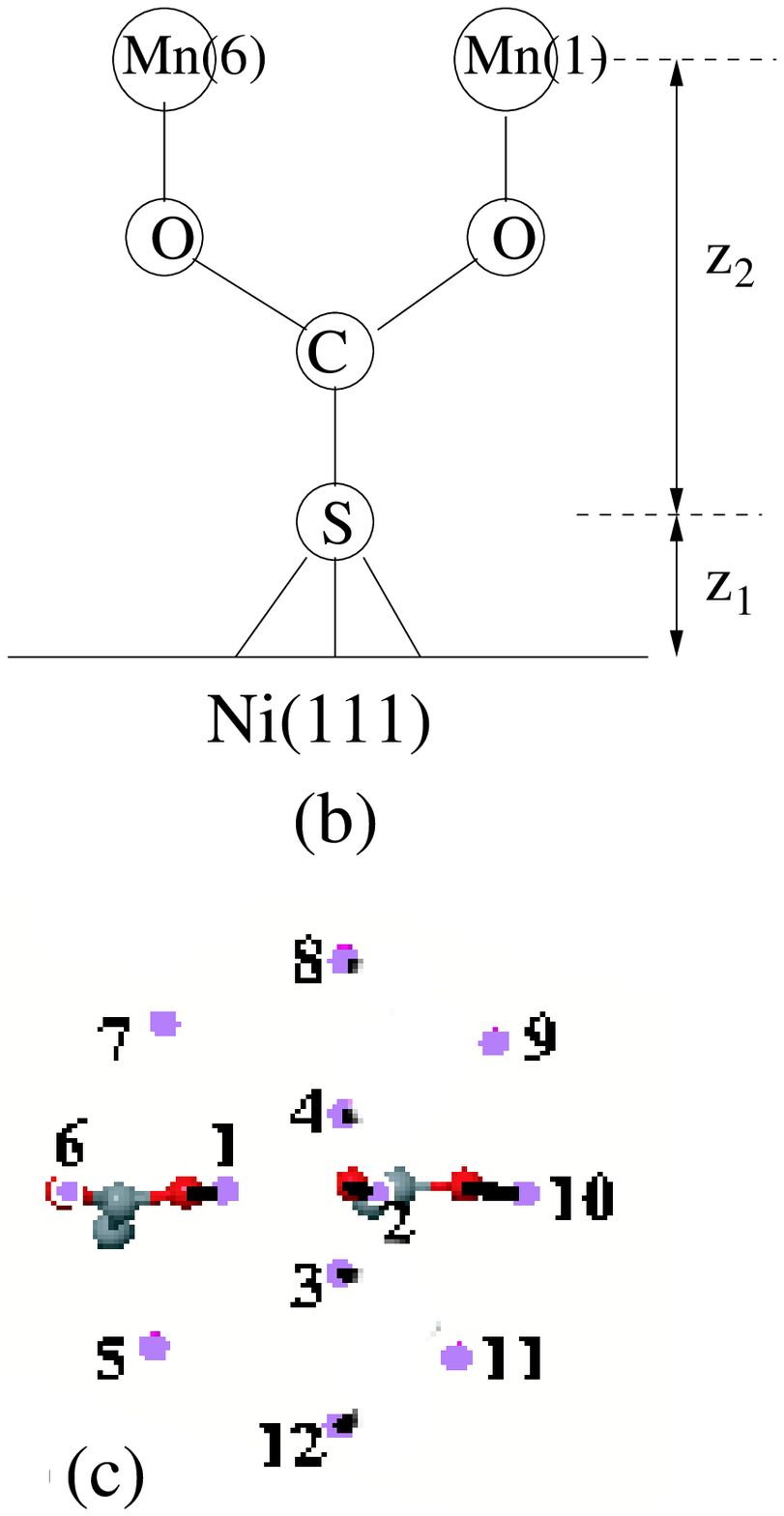}
\caption{(Color online) (a) Side view of the whole structure: a SMM Mn$_{12}$ adsorbed on a FM Ni(111) 
slab of six atomic layers, via two S atoms. (b) Part of the interface region. The vertical distances, 
$z_1$ and $z_2$, are listed in Table~\ref{table:3}. (c) Top view (in the $xy$ plane) showing the twelve 
Mn ions labeled, the O bonded to the Mn(1), Mn(2), Mn(6), and Mn(10) sites, and the C bonded to the O 
and S. Modified from Ref.[\onlinecite{PARK10}].}
\label{fig:geo}
\end{figure}

We consider a structure whose unit cell is shown in Fig.~\ref{fig:geo}(a). In the unit cell, a single Mn$_{12}$ 
molecule is adsorbed on a FM Ni(111) slab of six atomic layers, via two S atoms. This structure is referred to 
as `whole structure.'  
We separately optimize geometries of a bare Ni slab (without a Mn$_{12}$ molecule) and an isolated Mn$_{12}$ 
molecule, combining the optimized geometries to construct the whole structure.


Let us discuss the geometry and properties of a bare Ni slab. We find that the equilibrium lattice constant of 
bulk face-centered-cubic (fcc) Ni equals 3.500~\AA (3.522~\AA) using {\tt SIESTA} ({\tt VASP}), which agrees with 
experiments \cite{TAYL50}. With the lattice constant, a Ni(111) slab of $1 \times 1 \times 6$ atoms is constructed, 
and its interlayer separations are optimized with in-plane separations fixed, until the maximum force is less than 
0.03 eV/\AA~for {\tt SIESTA} (0.02~eV/\AA~for {\tt VASP}). The optimized interlayer separations are listed
in Table~\ref{table:zdist}. The spin moments of the six atomic layers are, from
the topmost to the bottommost layers, found to be (0.74, 0.68, 0.64, 0.64, 0.68, 0.74) $\mu_{\rm B}$,
respectively, using {\tt SIESTA}, while they are equal to (0.68, 0.68, 0.65, 0.65, 0.69, 0.68) 
$\mu_{\rm B}$, respectively, using {\tt VASP}. 


We briefly review the geometry and properties of an isolated Mn$_{12}$ molecule. More details can be found in 
Ref.[\onlinecite{SALV07}]. A standard Mn$_{12}$ molecule, [Mn$_{12}$O$_{12}$(CH$_3$COO)$_{16}$(H$_2$O)$_4$],
has the total magnetic moment of 20 $\mu_{\rm B}$ in the ground state, where the eight outer Mn spins 
(each Mn$^{3+}$ having 4~$\mu_{\rm B}$) align antiparallel to the four inner Mn spins (each Mn$^{4+}$ having 
3~$\mu_{\rm B}$). For a standard Mn$_{12}$ molecule, both the lowest unoccupied molecular orbital (LUMO) and 
the highest occupied molecular orbital (HOMO) arise from the Mn $d$ orbitals. To reduce computational cost, 
we simplify a standard SMM Mn$_{12}$ into [Mn$_{12}$O$_{12}$(HCOO)$_{16}$], where the CH$_3$ groups in the 
standard Mn$_{12}$ are replaced by H \cite{PEDE99}, and the four water molecules are removed \cite{SALV07}. 
This simplification does not affect the magnetic core of the Mn$_{12}$. We substitute S atoms for the two H 
atoms closest to the Ni surface [Fig.~\ref{fig:geo}(b)], such that the Mn$_{12}$ molecule bonds to the Ni 
surface via the S. The bond length between the C and S equals 1.88~\AA~(Fig.~\ref{fig:geo}(b), Table~\ref{table:3}). 
This slightly modified form of Mn$_{12}$ is referred to as a S-terminated Mn$_{12}$. For a S-terminated 
Mn$_{12}$ molecule, the magnetic moments of the Mn ions remain unchanged, but each S atom has the magnetic 
moment of 1~$\mu_{\rm B}$ aligned antiparallel to the net magnetic moment of the twelve Mn ions. As a result, 
a S-terminated Mn$_{12}$ molecule has the total magnetic moment of 18~$\mu_{\rm B}$ in the ground state. 
The magnetic moments listed in Table~\ref{table:1} are computed by placing a sphere around each atom with
the radius given in Table~\ref{table:WS}. (The radii for O and S used in this work differ from those in 
Ref.[\onlinecite{SALV07}].) The atomically resolved magnetic moments (Table~\ref{table:1}) are lower than 
the actual values because contributions from inter-atomic regions are not included. 
For a S-terminated Mn$_{12}$, the LUMO and HOMO originate from the S $p$ orbitals. 


We now discuss the geometry of a unit cell of the whole structure. We determine a unit cell in order for
the following conditions to be satisfied: (i) one Mn$_{12}$ molecule must fit the Ni surface area, (ii) 
neighboring Mn$_{12}$ molecules interact weakly, and (iii) the total number of basis sets per unit cell 
should be computationally feasible. A monoclinic unit cell of 
$17.33 \times 17.33 \times 36.00$~\AA$^3$ ($17.43 \times 17.43 \times 32.00$~\AA$^3$) 
suffices to meet the above conditions for {\tt SIESTA} ({\tt VASP}). The unit cell [Fig.~\ref{fig:geo}(a)]
consists of a Mn$_{12}$ molecule attached to a Ni slab of $7 \times 7 \times 6$ atoms via two S atoms 
and a vacuum layer of 13.72~\AA~(9.69~\AA)~for {\tt SIESTA} ({\tt VASP}). The two S atoms bond to 
hollow sites of the Ni(111) surface and to the C atoms close to the Mn(1), Mn(2), Mn(6), 
and Mn(10) sites [Figs.~\ref{fig:geo}(b),(c)]. The whole structure contains the total of 382 atoms, 
equivalent to 3450 valence electrons or 5552 orbitals. We sample $3 \times 3 \times 1$ $k$-points for 
our DFT calculations. Considering the large system size, full geometry relaxations are difficult to
achieve. Instead, we carry out partial geometry relaxations using {\tt SIESTA}, where the Mn$_{12}$, 
S, and Ni surface layer relax with the five Ni layers fixed. We find that the partial geometry relaxations 
do not alter our findings discussed below. We thus, henceforth, consider the unrelaxed whole structure 
only. For the unrelaxed geometry, our calculations show that large forces (the $z$ components only) 
act mainly on the two S atoms and the six Ni surface atoms bonded to the S. They are, respectively, 
$-$2.35, $-$2.35, 1.58, 1.27, 1.39, 1.44, 1.54, and 1.68~eV/\AA,~using {\tt SIESTA} 
($-$1.52, $-$1.52, 0.93, 0.70, 0.69, 0.73, 0.74, and 0.86~eV/\AA,~using {\tt VASP}). 

\section{Results and Discussion}

\subsection{Magnetic coupling type and strength}

We consider two spin configurations in the whole structure: (i) a FM configuration where the magnetic moment
of the Mn$_{12}$ is parallel to the magnetization of the Ni slab, and (ii) an AFM configuration where the
magnetic moment of the Mn$_{12}$ is now reversed such that it is antiparallel to the magnetization of the Ni slab.
Our DFT calculations show that the AFM configuration has a lower energy than the FM one, and that the energy 
difference between the two configurations equals 39.5 meV (51.6 meV), using {\tt SIESTA} ({\tt VASP}). 
The exchange-correlation potential within the PBE GGA does not fully treat self-interaction corrections of the 
localized $d$ states in the Coulomb potential. As a result, residual self-interactions induce slightly more 
diffuse $d$ electrons, which entails overestimated exchange coupling. To examine the effect of 
self-interaction corrections, we apply a DFT$+U$ method \cite{ANIS97} to the two spin configurations with 
$U$=4~eV for the Mn $d$ orbitals. Our DFT+$U$ calculations using {\tt VASP} show that the AFM configuration 
has still a lower energy than the FM, and that the energy difference equals 46.3~meV. This energy difference
does not differ much from that using DFT (Table~\ref{table:2}), so that we, henceforth, discuss results obtained
using DFT, unless specified otherwise.

\subsection{Electronic properties of the whole structure}

\begin{figure}
\includegraphics[width=8.cm, height=6.cm]{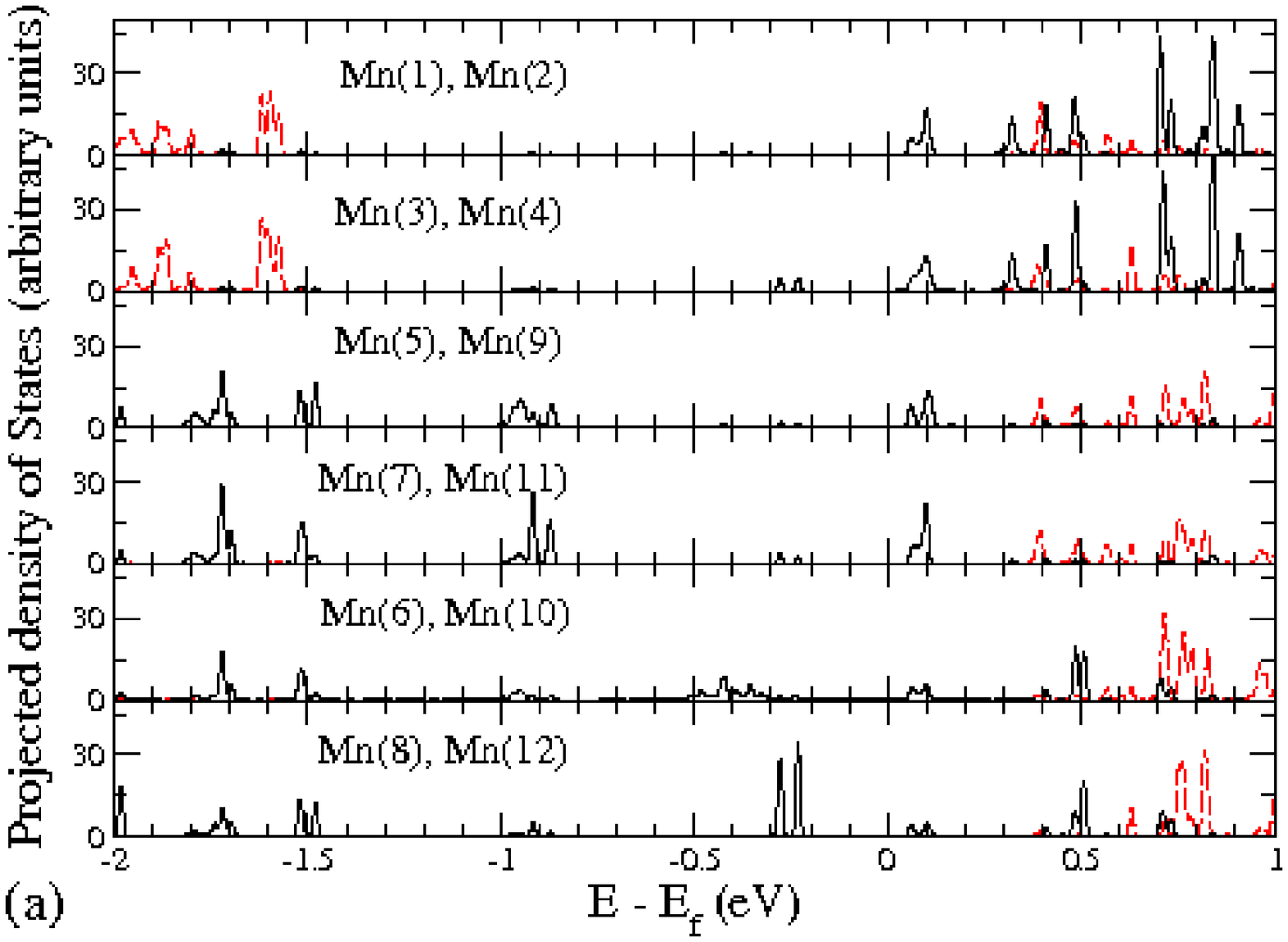} 
\hspace*{0.2cm}
\includegraphics[width=8.cm, height=6.cm]{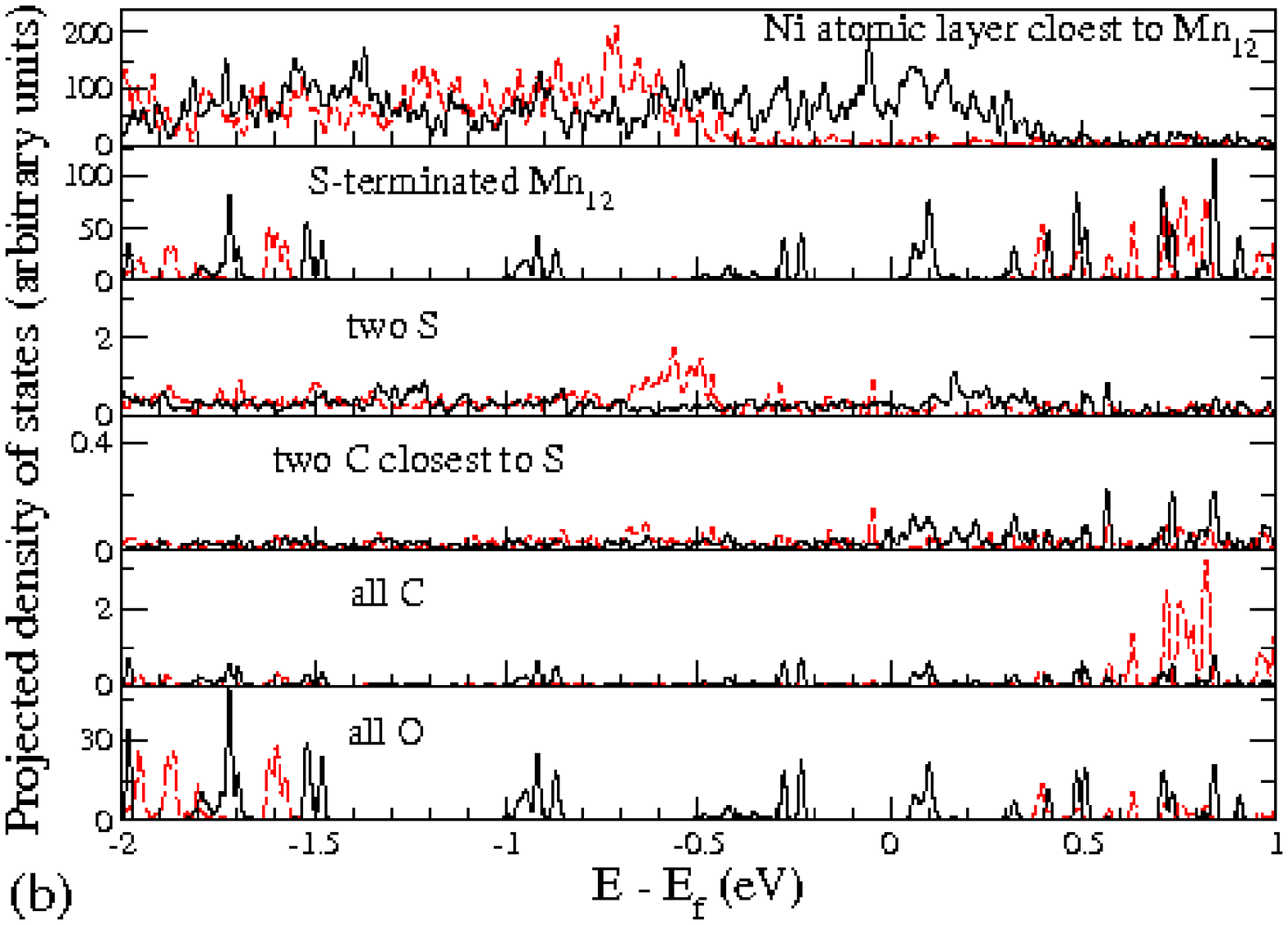}
\caption{(Color online) Spin-polarized DOS projected onto (a) the Mn $d$ orbitals, and (b) the Ni surface 
$d$ orbitals, all orbitals of the S-terminated Mn$_{12}$, the $p$ orbitals of the O, C, and S atoms, and 
the C $p$ orbitals bonded to the S atoms. The DOS is calculated for the AFM configuration. The majority 
(minority) spin is denoted as red dashed (black solid) curves. Notice the different vertical scales in (b).}
\label{fig:PDOS}
\end{figure}

We investigate the effect of the Ni slab on the electronic properties of the S-terminated Mn$_{12}$ in the AFM 
configuration. Using a Methfessel-Paxton \cite{METH89} smearing parameter $\sigma$=0.01 eV in {\tt VASP}, 
we calculate the density of states (DOS) projected onto the Mn $d$ orbitals, the Ni $d$ orbitals, and the $p$ 
orbitals of the O, S, and C atoms. As shown in Fig.~\ref{fig:PDOS}(a), the Mn $d$ orbitals broaden due to 
interactions with the Ni slab. Large broadening occurs for the $d$ orbitals of the Mn(6) and Mn(10) sites 
near 0.5~eV below the Fermi level, $E_{\rm f}$. Notice that the Mn(6) and Mn(10) sites are closest to the 
Ni surface [Fig.~\ref{fig:geo}(c)]. The O $p$ orbitals strongly hybridize with the Mn $d$ orbitals 
(Fig.~\ref{fig:PDOS}). In the band structure of Ni [top panel of Fig.~\ref{fig:PDOS}(b)], the minority-spin 
$d$ bands intersect $E_f$, while the majority-spin $d$ bands are completely filled. The S $p$ orbitals and 
the C $p$ orbitals bonded to the S atoms are delocalized due to the interactions with the Ni slab [third 
and fourth panels of Fig.~\ref{fig:PDOS}(b)]. The S $p$ orbitals are spin polarized, and strongly hybridize 
with the surface Ni $d$ orbitals than the C $p$ orbitals do.

\begin{figure}
\includegraphics[width=13.5cm, height=9.0cm]{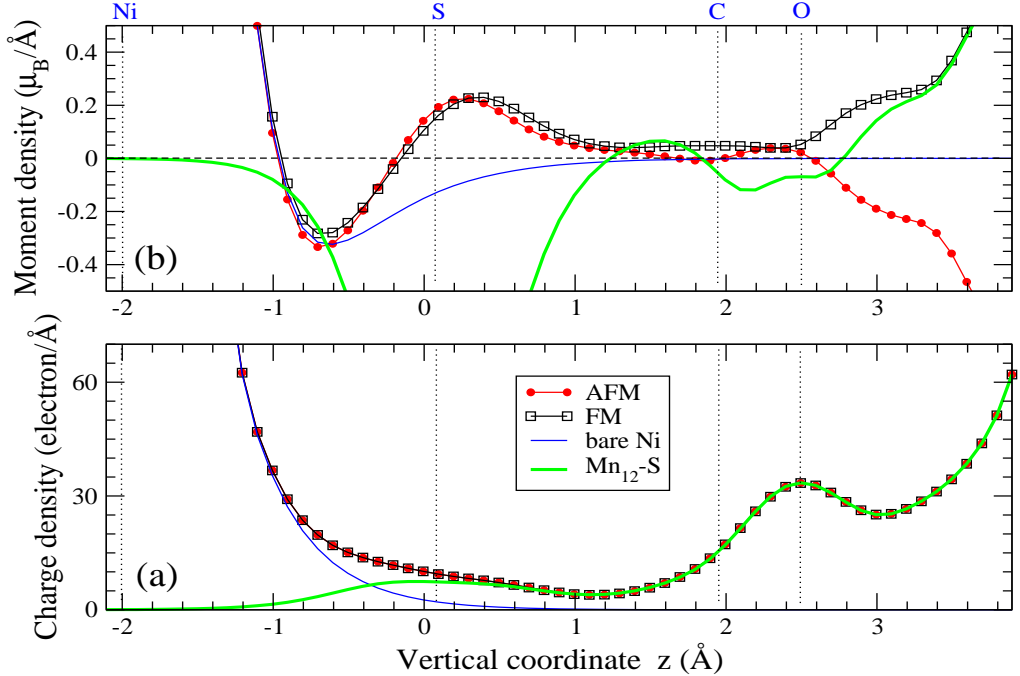}
\caption{(Color online) (a) Charge density and (b) magnetic moment density integrated over the
$xy$ plane, as a function of $z$ for the AFM (red filled circles) and FM (black empty squares) 
configurations, the bare Ni slab (blue thin solid curve), and the isolated S-terminated Mn$_{12}$ 
(green thick solid curve). The $z$ coordinates of the Ni surface layer, S, C, and O sites are 
marked. The Mn sites locate out of the range. The S atoms are located at $z=0.09$~\AA,
as in Ref.[\onlinecite{SALV07}].}
\label{fig:chgmag}
\end{figure}

\begin{figure}
\includegraphics[width=13.5cm, height=9.0cm]{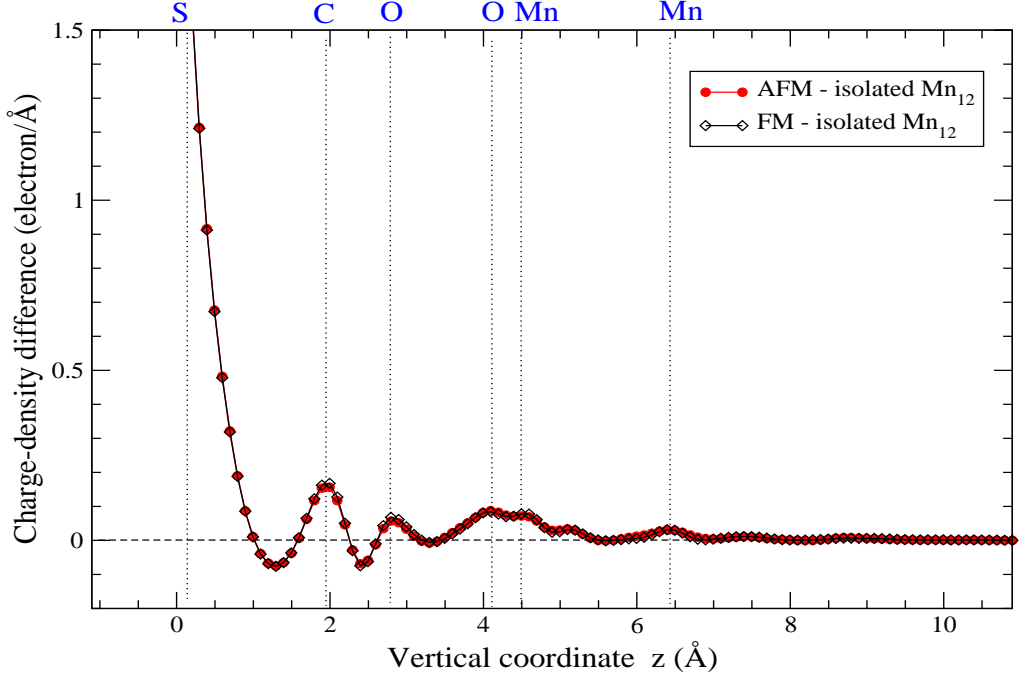}
\caption{(Color online) The difference in charge density between the whole structure and the isolated 
S-terminated Mn$_{12}$, $\Delta \rho(z)$, as a function of $z$ 
for the AFM (red filled circles) and FM (black empty diamonds) configurations. The $z$ coordinates of 
the S, the C and O closest to the S, and the Mn, are marked.}
\label{fig:charge}
\end{figure}

Let us now discuss the charge distribution of the whole structure as well as calculate charge transfer 
between the Ni slab and the S-terminated Mn$_{12}$ for the FM and AFM configurations. We perform our calculations 
using {\tt VASP}. As shown in Fig.~\ref{fig:chgmag}(a), the charge distribution of the bare Ni slab has a long 
tail which is deeply penetrated into the region beyond the S. The charge distribution of the whole structure 
coincides with that of the isolated Mn$_{12}$ near the mid-distance between the S and the C [Fig.~\ref{fig:chgmag}(a)]. 
To compute the charge transfer, we integrate over the $xy$ plane the charge density of the whole structure, 
$\rho_{\rm whole}(x,y,z)$, and the charge density of the isolated S-terminated Mn$_{12}$, $\rho_{\rm SMM,isol}(x,y,z)$. 
We then take a difference between the two integrated charge densities: 
$\Delta \rho(z) = \int_{xy} (\rho_{\rm whole} - \rho_{\rm SMM,isol}) dx dy$. The charge-density difference, 
$\Delta \rho(z)$ (Fig.~\ref{fig:charge}), peaks at the linker molecules, the S, into which the charge distribution 
of the Ni slab is penetrated [Fig.~\ref{fig:chgmag}(a)]. We calculate the charge transfer by an integration 
of $\Delta \rho(z)$ over $z$. We find that some amount of charge is transferred from the Ni slab to the 
Mn$_{12}$ molecule. The magnitude of the charge transfer is sensitive to the lower bound of the integration
because the exact boundary between the Ni slab and the Mn$_{12}$ is hard to be determined in the whole structure. 
For instance, when we integrate $\Delta \rho(z)$ from $z=-0.94$~\AA~to $z=11.90$~\AA,
~the charge transferred equals 13.77 (13.78) electrons for the AFM (FM) configuration.  
Here $z=-0.94$~\AA~represents the mid-distance between the Ni surface layer and the S, and 
the S atoms are located at $z=0.09$~\AA,~as in Ref.[\onlinecite{SALV07}].

\subsection{Magnetic properties of the whole structure}

\begin{figure}
\includegraphics[width=13.5cm, height=9.0cm]{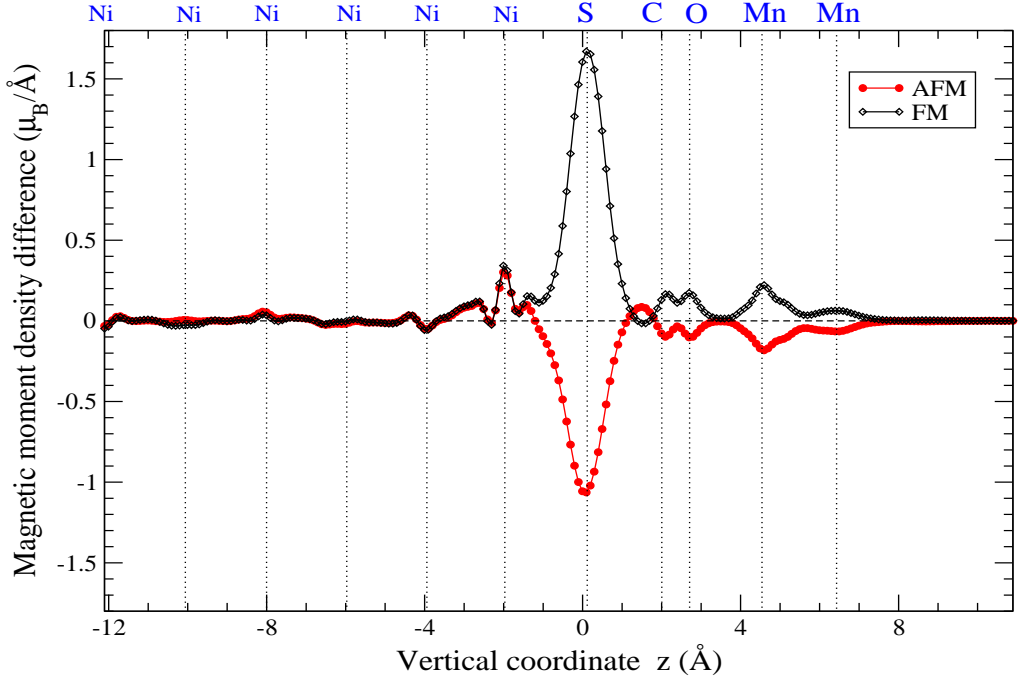}
\caption{(Color online) The difference in magnetic-moment density between the whole structure and the sum of 
the bare Ni slab and the isolated S-terminated Mn$_{12}$, $\Delta \rho_M(z)$, as a function of $z$ 
for the AFM (red filled circles) and FM (black empty diamonds) configurations. The $z$ coordinates 
of the six Ni atomic layers, the S, the C and O closest to the S, and the Mn, are marked.}
\label{fig:magmom}
\end{figure}

\begin{figure}
\includegraphics[width=13.5cm, height=6.cm]{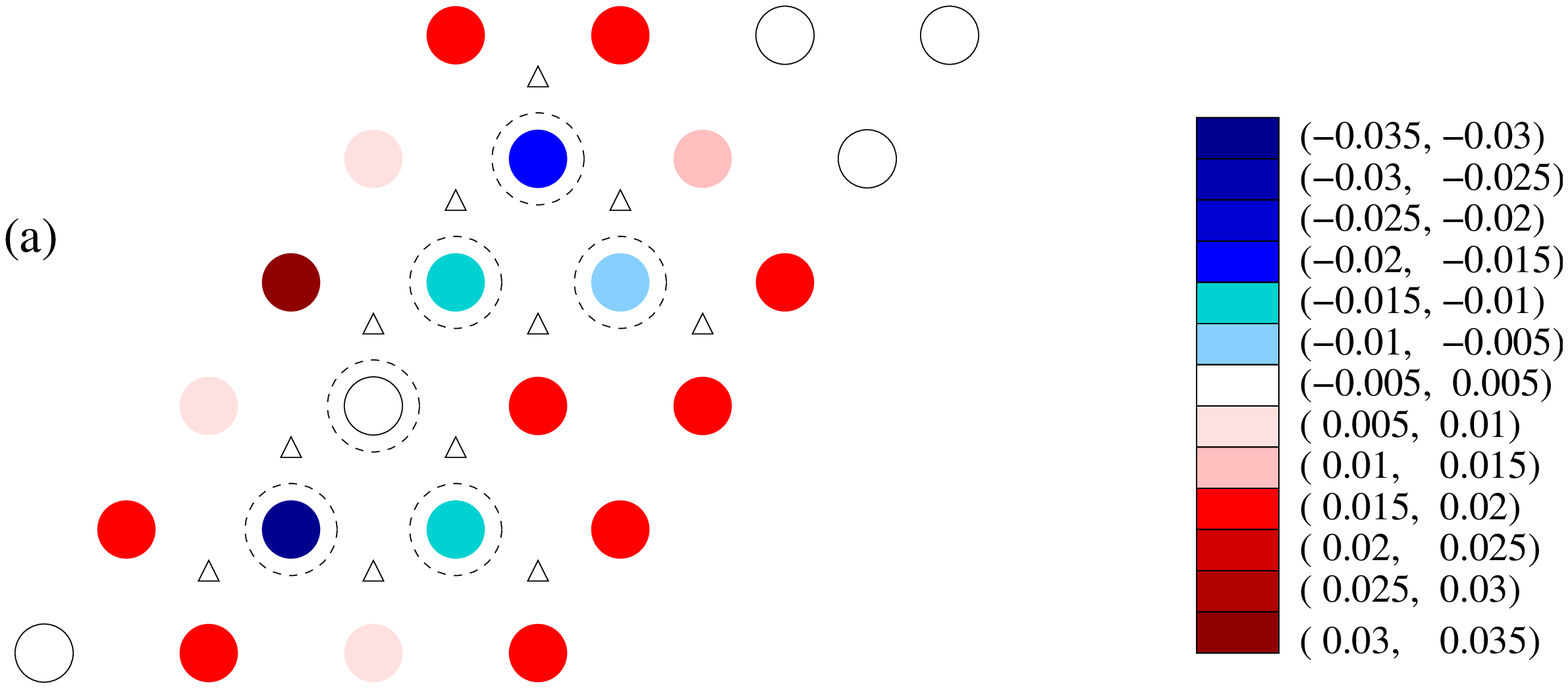}
\hspace{6.0truecm}
\includegraphics[width=9.cm, height=6.cm]{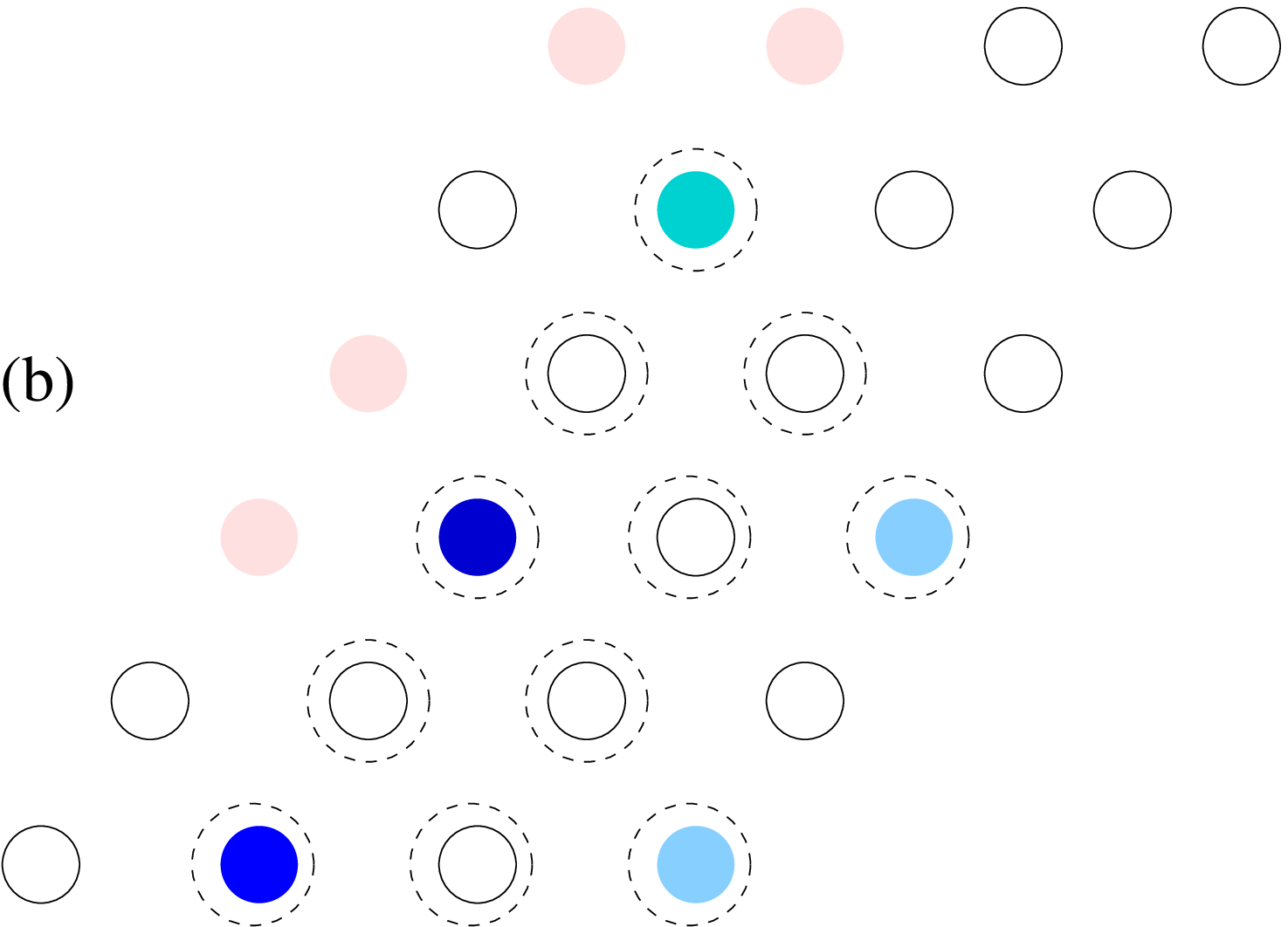}
\caption{(Color online) Magnetic moments induced in (a) the topmost Ni surface layer and (b)
the second Ni layer from the surface in units of $\mu_{\rm B}$ for the AFM configuration.
Not all of the Ni sites per atomic layer are shown.
The dashed circles in (a) represent the sites where the S ions are bonded. 
The sites marked by the triangles in (a) correspond to the dashed circles in (b) in the
$xy$ plane.}
\label{fig:Nimag}
\end{figure}

We examine how the adsorption modifies the magnetic moments of the S-terminated Mn$_{12}$ and of the Ni slab 
as a function of $z$ for the FM and AFM configurations, using {\tt VASP}. 
Similarly to the charge-transfer calculation, we integrate 
over the $xy$ plane the spin density of the whole structure, $\rho_{\rm whole}^{\rm M}$, the spin density of the 
isolated S-terminated Mn$_{12}$, $\rho_{\rm SMM,isol}^{\rm M}$, and the magnetization of the bare Ni slab, 
$\rho_{\rm Ni,bare}^{\rm M}$. We refer to these integrated quantities as one-dimensional magnetic-moment densities 
and magnetization depending on $z$. We consider the following difference in one-dimensional magnetic-moment density: 
$\Delta \rho_M(z)=\int_{xy} [\rho_{\rm whole}^{M} - (\int_{xy} \rho_{\rm Ni, bare}^{\rm M} \pm \int_{xy} \rho_{\rm SMM,isol}^{\rm M})] dx dy$,
where the positive (negative) sign in front of the third term is applied to the FM (AFM) configuration. As shown 
in Fig.~\ref{fig:magmom}, the most prominent effects of the adsorption in the difference $\Delta \rho_M(z)$
are (i) that the difference peaks at the S ions, and (ii) that the AFM configuration possesses a smaller peak 
amplitude in the difference than the FM configuration does. For the isolated Mn$_{12}$, each S ion has the 
magnetic moment of 1~$\mu_{\rm B}$, while for the whole structure, the S ions lose the spin polarization to great
extent. As a result, the peak in $\Delta \rho_M(z)$ occurs at the S ions. Notice that in the isolated Mn$_{12}$, 
the magnetic moment of each S ion aligns antiparallel to the net magnetic moment of the twelve Mn ions
[Fig.~\ref{fig:chgmag}(b)]. In the whole structure, however, the magnetic moment of each S ion aligns parallel 
to the magnetization of the Ni slab, independent of the sign of the magnetic coupling between the Mn$_{12}$ and 
the Ni slab (Table~\ref{table:1}). This entails that the peak amplitude for the AFM configuration is smaller than 
that for the FM configuration. (Effects (i) and (ii) are further discussed in Sec.IV.D.)  
For the Ni slab, the magnetic moments of the top two layers in proximity to the Mn$_{12}$ substantially change 
due to the interactions with the Mn$_{12}$. As shown in Fig.~\ref{fig:Nimag}, great changes appear near the  
sites where the S ions are bonded. The spatial dependence of the magnetic moments induced in the Ni slab for the
FM configuration is similar to that for the AFM illustrated in Fig.~\ref{fig:Nimag}. The main features discussed 
in this section remain unchanged with the application of the GGA+U method.

\begin{figure}
\includegraphics[width=13.5cm, height=9.0cm]{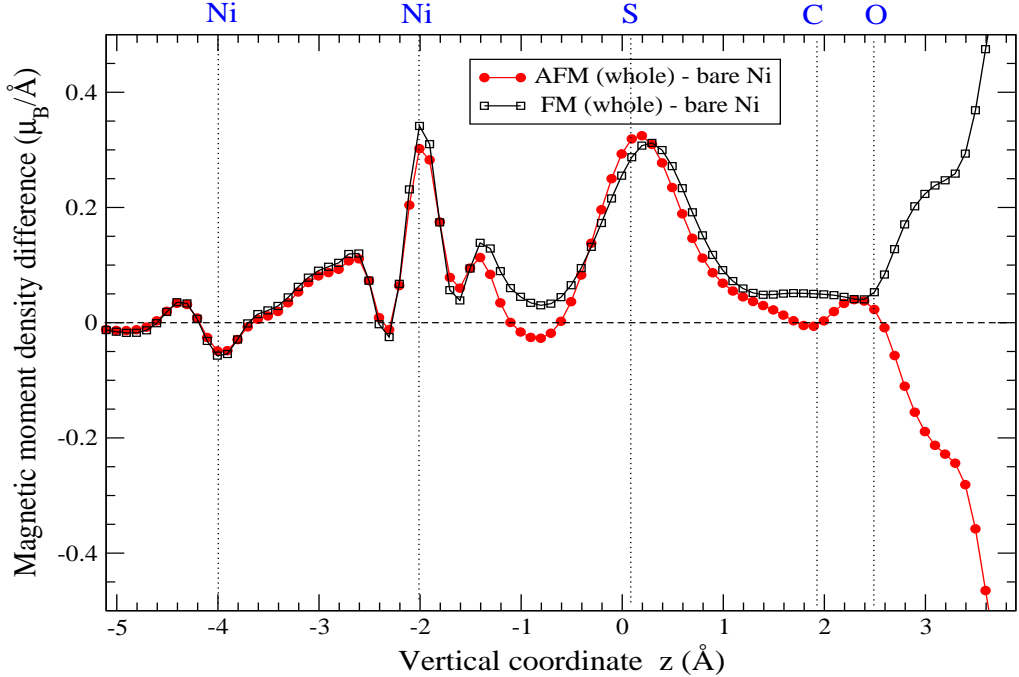}
\caption{(Color online) The difference in one-dimensional magnetic-moment density between the whole 
structure and the bare Ni slab, $\int_{xy} (\rho_{\rm whole}^{\rm M} - \rho_{\rm Ni,bare}^{\rm M}) dx dy$, 
as a function of $z$ for the AFM (red filled circles) and FM (black empty squares) configurations.}
\label{fig:diff}
\end{figure}

Let us now discuss a way to calculate the total magnetic moment induced in the Ni slab, $\Delta M_{\rm Ni}^{\rm ind}$,
and the magnetic moment of the adsorbed Mn$_{12}$, $M_{\rm SMM}^{\rm ads}$, within the GGA and GGA+U formalisms, 
using {\tt VASP}. The difference in magnetic moment between the whole structure and the bare Ni slab, 
[$M_{\rm whole} - M_{\rm Ni}^{\rm bare}$], has contributions from both $\Delta M_{\rm Ni}^{\rm ind}$ and
$M_{\rm SMM}^{\rm ads}$, so that $[M_{\rm whole} - M_{\rm Ni}^{\rm bare}=M_{\rm SMM}^{\rm ads} + \Delta M_{\rm Ni}^{\rm ind}]$. 
We first compute [$M_{\rm whole} - M_{\rm Ni}^{\rm bare}$] including contributions from inter-atomic regions. 
To find the value of $\Delta M_{\rm Ni}^{\rm ind}$, we integrate the difference in one-dimensional magnetic-moment 
{\it density} between the whole structure and the bare Ni slab, 
$\int_{xy} (\rho_{\rm whole}^{\rm M} - \rho_{\rm Ni,bare}^{\rm M}) dx dy$ (Fig.~\ref{fig:diff}), from the $z$ 
coordinate of the bottommost Ni layer (farthest from the surface layer, $z=-12.10$~\AA)~to the mid-distance between 
the Ni surface layer and the S atoms ($z=-0.94$~\AA).~
Similarly to the charge-transfer calculation, the value of $\Delta M_{\rm Ni}^{\rm ind}$ depends on the upper bound 
in the integration, and the difference in $\Delta M_{\rm Ni}^{\rm ind}$ between the AFM and FM configurations 
is very small (Fig.~\ref{fig:diff}). Then using the calculated values of 
[$M_{\rm whole} - M_{\rm Ni}^{\rm bare}$] and $\Delta M_{\rm Ni}^{\rm ind}$, we find the value of $M_{\rm SMM}^{\rm ads}$. 

Within the GGA, the difference, [$M_{\rm whole} - M_{\rm Ni}^{\rm bare}$], amounts to
$-$19.29 (20.54)~$\mu_{\rm B}$ for the AFM (FM) configuration. When we integrate the one-dimensional 
density difference from $z=-12.10$ to $z=-0.94$~\AA,~we find that $\Delta M_{\rm Ni}^{\rm ind}$ amounts to 
0.24~$\mu_{\rm B}$ (0.23~$\mu_{\rm B}$) for the AFM (FM) configuration (Fig.~\ref{fig:diff}). Thus, 
$M_{\rm SMM}^{\rm ads}$ equals $-$19.53~(20.31)~$\mu_{\rm B}$ for the AFM (FM) configuration. Within the
GGA+U formalism, the difference, [$M_{\rm whole} - M_{\rm Ni}^{\rm bare}$] reaches $-$19.05~(20.46)~$\mu_{\rm B}$ 
for the AFM (FM) configuration. Since the $U$ term is added to the Mn $d$ orbitals only, the total magnetic moment
induced in the Ni slab would not change with the addition of the $U$ term. Thus,
$M_{\rm SMM}^{\rm ads}$ equals $-$19.29~(20.23)~$\mu_{\rm B}$ for the AFM (FM) configuration. Considering that
the magnetic moment of an isolated S-terminated Mn$_{12}$ equals 18~$\mu_B$, induced magnetic moments on the
adsorbed Mn$_{12}$ are shown in Table~\ref{table:Mn12}.

To summarize the above calculations, when a S-terminated Mn$_{12}$ molecule is adsorbed on a FM metal such as Ni, both 
the GGA and GGA+U calculations suggest that the adsorbed Mn$_{12}$ molecule has the magnetic moment deviated 
from 20~$\mu_B$, and that the sign of its deviation from 20~$\mu_B$ depends on the sign of the magnetic coupling between the 
Mn$_{12}$ and the FM substrate (Table~\ref{table:Mn12}). This result differs from the case that a S-terminated 
Mn$_{12}$ molecule is adsorbed on a non-magnetic metal such as Au. The magnetic moment of a S-terminated Mn$_{12}$ 
adsorbed on Au increases to 19.8~$\mu_{\rm B}$ \cite{SALV07}. Notice that an isolated standard 
Mn$_{12}$ molecule has the magnetic moment of 20~$\mu_B$. 

We now investigate the effect of the adsorption on the magnetic anisotropy of the AFM configuration. We perform 
DFT calculations including SOC using {\tt VASP}. Considering the large system size, we treat the SOC 
non-self-consistently in collinear cases. Previous theoretical calculations \cite{SALV07} have shown that a non-self-consistent 
treatment of SOC produces a slightly lower magnetic anisotropy barrier (lower by 14\% for an isolated 
standard Mn$_{12}$) than an experimental value or a self-consistently calculated value with a correct magnetic 
easy axis \cite{SALV07}. Our calculation reveals that a bare Ni(111) slab of six atomic layers possesses 
the magnetic easy axis parallel to the surface (which agrees with experiments \cite{OBRI94}), and that the 
magnetic anisotropy barrier of the slab ($7 \times 7 \times 6$ atoms) equals 66.9~meV. An isolated S-terminated 
Mn$_{12}$ molecule has, however, the magnetic easy axis normal to the surface, and its magnetic anisotropy barrier 
is found to be 4.88~meV. For the whole structure, we consider only two collinear spin configurations: (i) both the 
magnetic moment of the Mn$_{12}$ and the magnetization of the Ni slab are parallel to the Ni surface, and (ii) they 
are perpendicular to the surface. Our calculations show that the first collinear configuration provides a lower energy
than the second configuration, which implies 
that the Ni slab prevalently contributes to the magnetic anisotropy of the whole structure. The energy difference 
between the two collinear configurations equals 55.3~meV. This energy difference is lower than the magnetic 
anisotropy barrier of the bare Ni slab because for the given molecular orientation the magnetic easy axis of 
the isolated S-terminated Mn$_{12}$ is perpendicular to that of the bare Ni slab. A more quantitative analysis 
requires a self-consistent treatment of SOC and inclusion of noncollinear spin configurations and dipolar interactions, 
which is beyond the scope of the current study.

\subsection{Mechanisms of magnetic coupling}

To understand the mechanisms of the magnetic coupling between the Mn$_{12}$ and the Ni slab, we consider direct 
exchange coupling and superexchange coupling \cite{GOOD63}. In the whole structure, the shortest 
separation between the Mn ions and the Ni surface atoms (6.58~\AA) is too large for direct exchange coupling 
to be effective. Thus, the Ni $d$ orbitals do not hybridize with the Mn $d$ orbitals. As a result, superexchange 
and/or Ruderman-Kittel-Kasuya-Yosida (RKKY)-like interactions \cite{ZIMA72,HEWS93} may cause the magnetic coupling. 
In numerous systems, either superexchange or RKKY-like interactions play a role in magnetic coupling. 
In this work, we consider superexchange coupling exclusively because it explains the magnetic coupling of interest.

\begin{figure}
\includegraphics[width=5.2cm, height=3.cm]{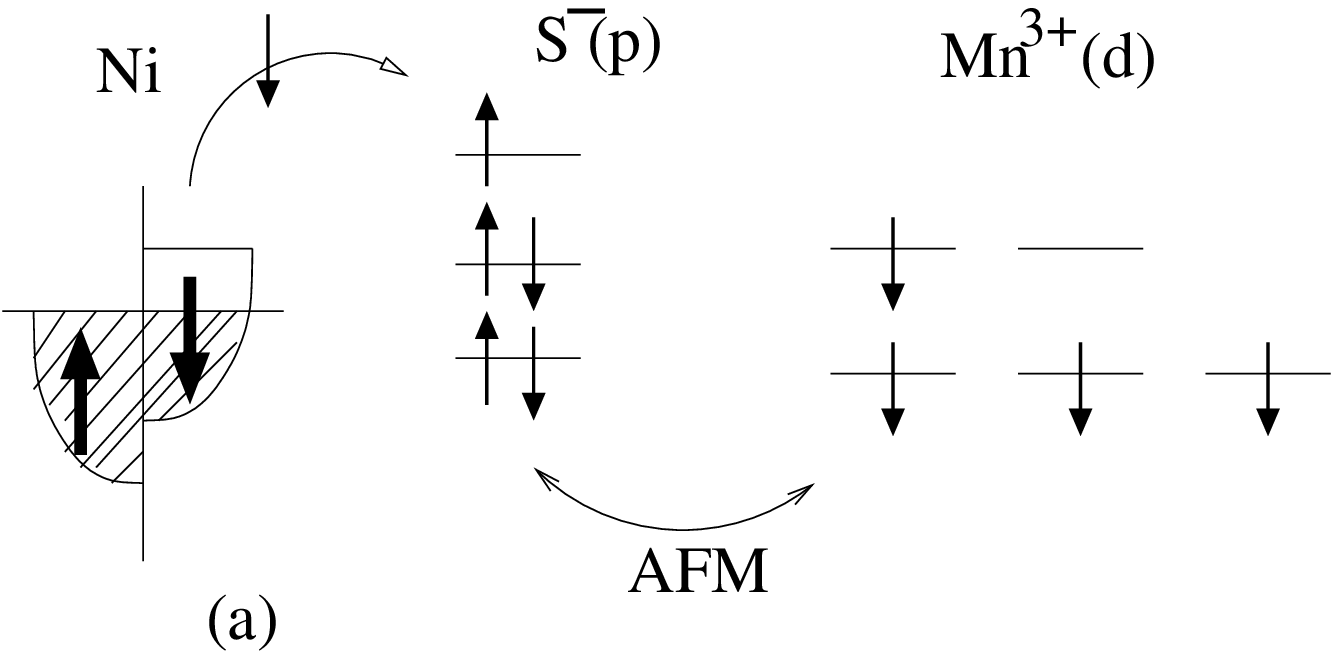}
\hspace{0.5cm}
\includegraphics[width=5.2cm, height=3.cm]{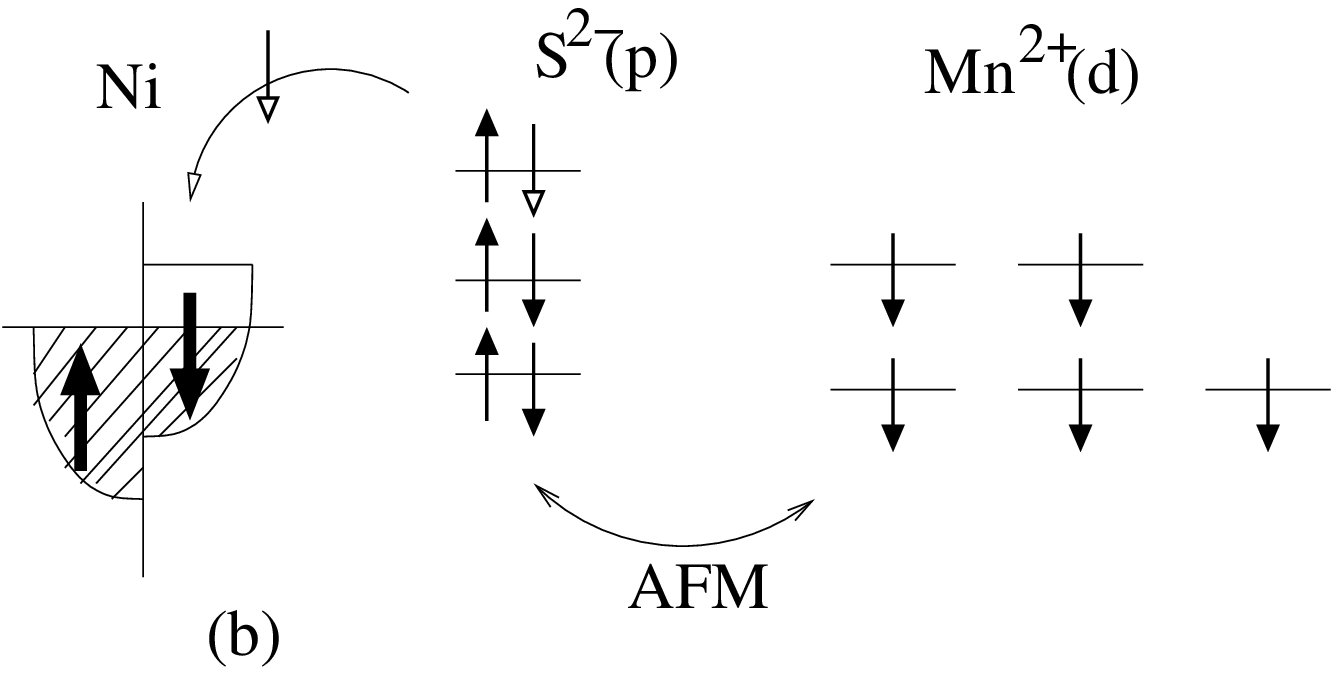}
\hspace{0.5cm}
\includegraphics[width=4.cm, height=2.9cm]{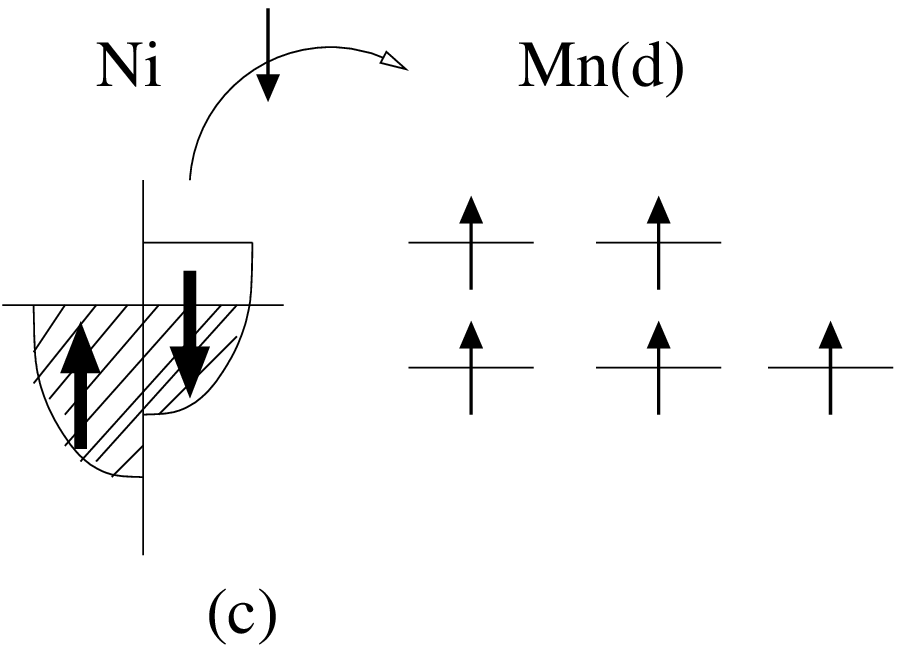}
\caption{Magnetic coupling via superexchange interactions for (a) the whole structure and (b) a single Mn atom 
adsorbed on Ni(111) via S. Magnetic coupling via hybridization for (c) a single Mn atom directly bonded to Ni(111). 
The majority and minority-spin Ni 3$d$ bands at zero temperature, S 2$p$ orbital levels (degenerate), and Mn 3$d$ 
orbital levels, are illustrated. The Mn $d$ levels for (b) and (c) are degenerate.}
\label{fig:coupling}
\end{figure}

Let us examine contributions of superexchange interactions to the magnetic coupling. In the whole structure, 
the shortest separation between the S sites and the Ni surface atoms is set to be 2.52~\AA~[Fig.~\ref{fig:geo}(b), 
Table~\ref{table:3}]. As discussed, charge transfer occurs from the Ni slab to the S-terminated Mn$_{12}$ molecule, 
or to the S ions which is the closest to the Ni slab [Fig.~\ref{fig:coupling}(a)]. The fact that the LUMO 
and HOMO of the Mn$_{12}$ arise from the S $p$ orbitals, corroborates the charge transfer to the S ions. 
The band structure of Ni entails favorable transfer of minority-spin electrons. Thus, the total magnetic
moment induced in the Ni slab, $\Delta M_{\rm Ni}^{\rm ind}$, is positive (Sec.IV.C), and the difference in
one-dimensional magnetic moment density between the whole structure and the bare Ni slab (Fig.~\ref{fig:diff})
has a positive peak amplitude at the Ni surface layer. This charge transfer lowers the magnetic moments of the 
S ions, resulting in the magnetic moments parallel to the magnetization of the Ni slab [Fig.~\ref{fig:coupling}(a)]. 
Let us now focus on the region between the S and the Mn ions [Fig.~\ref{fig:geo}(b)]. The magnetic moments of the 
C and O ions closest to the S ions are much smaller than those of the S ions (Table~\ref{table:1}). As shown in 
Fig.~\ref{fig:geo}(b), C and O anions are located between the S and the Mn(1) and Mn(6). The bond angle created by 
the Mn(6) [or Mn(1)] and the O and C bonded to the Mn [Fig.~\ref{fig:geo}(b)] equals 123 degrees. The bond angle 
of S-C-O in Fig.~\ref{fig:geo}(b) falls on 114-118 degrees. While the S ions have dominant contributions from 
2$p_x$ and 2$p_y$ orbitals, the C and O anions have equal contributions from 2$p_x$, 2$p_y$, and 2$p_z$ orbitals. 
The Mn(1) (Mn$^{4+}$) possesses exactly half-filled spin-polarized $t_{2g}$ orbitals. Our calculations suggest
that this inner Mn ion is ferromagnetically coupled to the magnetic moments of the S, C, and O ions. 
The Mn(6) (Mn$^{3+}$) has spin-polarized half-filled $t_{2g}$-like orbitals and quarter-filled $e_{g}$-like 
orbitals. Our calculations indicate that the outer Mn ion is antiferromagnetically coupled to the magnetic
moments of the S, C, and O ions. The same logic is applied to the other linker molecule.
Therefore, the magnetic moments of the S, C, and O ions are antiferromagnetically 
coupled with the net magnetic moment of the twelve Mn ions. As a result, the superexchange interactions render 
overall AFM coupling between the local magnetic moments of the Ni and the total magnetic moment of the Mn ions 
via the S, C, and O $p$ orbitals [Fig.~\ref{fig:coupling}(a)]. We emphasize that not only the separation between 
the Ni and the Mn ions but also the type and geometry of bonding to the Ni and Mn, play an important role for the 
superexchange interactions.

To gain further insight into the nature of the magnetic coupling, we investigate magnetic coupling in 
the following testbed systems: (i) a single Mn atom adsorbed on a Ni(111) slab via one S atom 
[Fig.~\ref{fig:coupling}(b)], and (ii) one Mn atom directly bonded to a Ni(111) slab [Fig.~\ref{fig:coupling}(c)]. 
The magnetic coupling in system (i) seems to have similar mechanisms to that in the whole structure, 
in that the magnetic moment of the Mn interacts with the Ni slab through the S, and yet system (i) is
much simpler than the whole structure. Now if we remove the S atom from system (i) and bring the Mn
atom closer to the Ni slab, the characteristics of the magnetic coupling are qualitatively modified. That 
is the case for system (ii). 

\begin{figure}
\includegraphics[width=13.5cm, height=9.0cm]{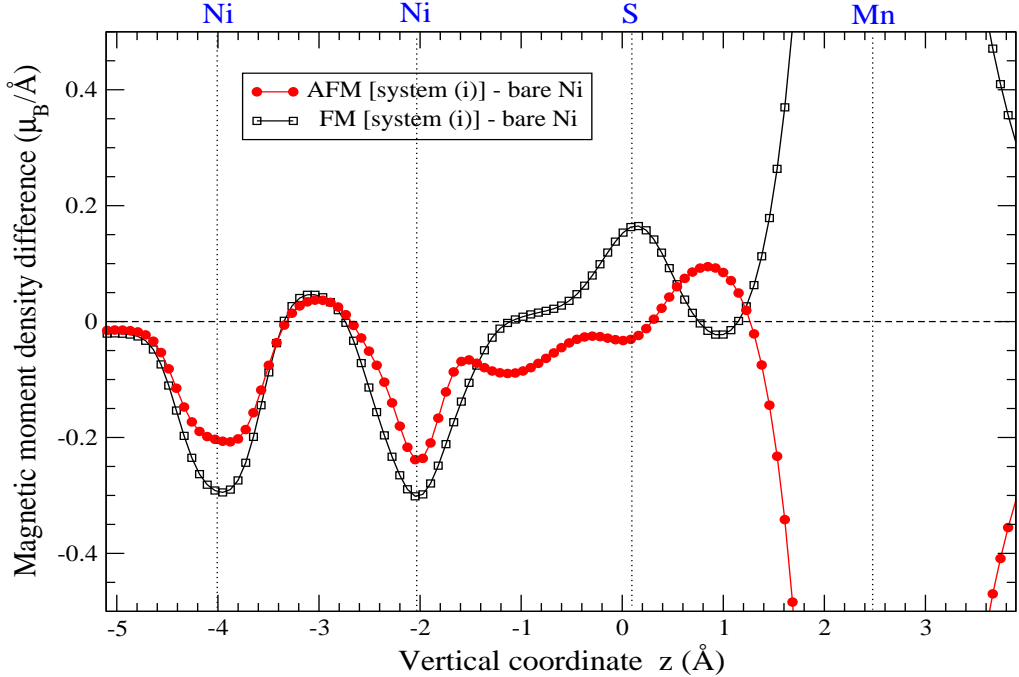}
\caption{(Color online) The difference in one-dimensional magnetic-moment density between system (i) 
and the bare Ni slab, $\int_{xy} (\rho_{\rm system \: (i)}^{\rm M} - \rho_{\rm Ni,bare}^{\rm M}) dx dy$, 
as a function of $z$ for the AFM (red filled circles) and FM (black empty squares) configurations. 
Calculated using {\tt VASP}.}
\label{fig:diff-b}
\end{figure}

Let us first discuss the magnetic coupling in system (i). We set the shortest distance between the S and Ni 
surface atoms to be the same as that for the whole structure (Table~\ref{table:3}), and assign positive spin
polarization to the Ni slab in both AFM and FM configurations. The S ion bonds to hollow sites of a Ni slab 
of $5 \times 5 \times 6$ atoms, and the bond angle of Mn-S-Ni equals 145.5$^{\circ}$. Our DFT calculations 
using {\tt SIESTA} suggest that similarly to the whole structure, the Mn ion favors AFM coupling to the Ni 
slab via the S anion (Table~\ref{table:3}). As in the whole structure,  
a major contribution to the magnetic coupling originates from a superexchange interaction via the S anion. 
An isolated Mn-S dimer has the ground-state spin of $S=5/2$, where the Mn ion carries most of the spin 
polarization. Since the S $p$ orbitals are completely filled and the minority-spin Ni $d$ bands are not 
fully filled, minority-spin electrons can be transferred now from the S to the Ni slab [Fig.~\ref{fig:coupling}(b)]. 
Consequently, the S $p$ orbitals possess slight spin polarization parallel to the magnetization of the Ni 
slab, and negative magnetic moments are induced in the Ni slab (Fig.~\ref{fig:diff-b}). The S $p$ orbitals 
are antiferromagnetically coupled to the Mn $d$ orbitals. Overall, this leads to AFM coupling between the 
Mn and the Ni (Fig.~\ref{fig:coupling}(b), Table~\ref{table:3}). The magnetic coupling in system (i) is 
stronger than that in the whole structure due to the following reasons: (1) The vertical separation between 
the Mn and the Ni surface atoms in system (i) is shorter than that in the whole structure (Table~\ref{table:3}). 
(2) A superexchange interaction through S is stronger than that via O or C, since S atoms more diffuse. In
the whole structure, the superexchange interactions are weakened by pathways through the S, C, and O
orbitals. For the given geometry of system (i), whether the 3$d$ orbital levels of the adatom are less 
than or more than half-filled, would not change the sign of the magnetic coupling because the magnetic 
coupling is mainly governed by a superexchange interaction. 

We now examine the magnetic coupling in system (ii). In this system, the Mn atom directly bonds to hollow 
sites of a Ni slab of $5 \times 5 \times 6$ atoms, and the Mn $d$ orbital levels are exactly half-filled 
[Fig.~\ref{fig:coupling}(c)]. Our DFT calculations using {\tt SIESTA} show that stable FM coupling is formed 
between the Mn and the Ni (Table~\ref{table:3}), in contrast to system (i) and the whole structure. 
In system (ii), the bond length between the Mn and the Ni is short (Table~\ref{table:3}), and so the 
minority-spin Mn $d$ orbitals hybridize with the minority-spin Ni $d$ orbitals. This results in FM coupling 
between the Mn and the Ni, which agrees with other theoretical calculations \cite{LOUN05}.
In system (ii), if the 3$d$ orbital levels of the adatom are less than half-filled, then hybridization 
favors AFM coupling between the adatom and the Ni \cite{LOUN05}.

\section{Conclusion}

We have investigated magnetic coupling between the SMM Mn$_{12}$ and a FM Ni(111) substrate, using DFT and a DFT+U method.
Our DFT and DFT+U calculations have shown that the Mn$_{12}$ interacts antiferromagnetically with the Ni substrate via S, 
and that the magnetic moment of the Mn$_{12}$ is deviated from that of an isolated standard Mn$_{12}$, 
due to interactions with the Ni substrate. We have also found that the magnetic anisotropy of the whole structure is 
dictated by the anisotropy of the Ni slab, and that the magnetic anisotropy barrier of the whole structure depends 
on the orientation of the Mn$_{12}$ relative to the Ni slab. The magnetic coupling is mainly caused by superexchange 
interactions between the magnetic moments of the Ni and Mn via the S, C, and O anions. Superexchange interactions take 
into account the bonding types and bonding geometries of linker molecules to the Ni slab and the Mn$_{12}$.  
Our findings can be observed in XMCD experiments and provide insight into magnetic coupling of other types of SMM 
to FM substrates.

\begin{acknowledgments}
K.P. was supported by NSF DMR-0804665 and the Jeffress Memorial Trust Funds. The author thanks Salvador Barraza-Lopez and 
Jaime Ferrer for discussions and reading the manuscript, and thanks Jean J. Heremans for discussions. Computational support 
was provided by the SGI Altix Linux Supercluster (Cobalt) and Intel 64 Cluster (Abe) at the National Center for Supercomputing
Applications under DMR060011, and by Virginia Tech Linux clusters and Advanced Research Computing.
\end{acknowledgments}

\clearpage

\begin{table}
\begin{center}
\caption{Vertical distances $z$ (in \AA),~bond lengths $d$ (in \AA),~and an energy difference
$\Delta E$ (in eV) between the AFM and FM configurations. $\Delta E < 0$ ($\Delta E > 0)$ implies 
stable AFM (FM) coupling. For the calculations of $\Delta E$ using {\tt SIESTA}, we sample 
$3 \times 3 \times 1$ $k$-points for the whole structure and $4 \times 4 \times 1$ $k$-points 
for systems (i) and (ii).}
\label{table:3}
\begin{ruledtabular}
\begin{tabular}{c|c|c|c|c|c|c}
       & $z_1=z$(Ni-S)    & $z_2=z$(S-Mn)    &  $d$(Ni-Mn)  &  $d$(Ni-S)  & $d$(S-C)   & $\Delta E$ \\ \hline
whole structure & 2.08     & 4.48     &  6.58       &  2.52       & 1.88       &  $-$0.0395      \\
system (i)       & 2.08     & 2.40     &  4.70       &  2.52       & N/A        &  $-$0.1050      \\
system (ii)       & N/A      & N/A      &  2.80       & N/A         & N/A        &  0.1692     
\end{tabular}
\end{ruledtabular}
\end{center}
\end{table}

\begin{table}
\begin{center}
\caption{Optimized interlayer separations (in \AA)~for the Ni slab of six atomic layers.
$\Delta z_{ij}$ indicates the vertical distance between the $i$th and the $j$th layers. $a$ is the equilibrium
lattice constant for bulk fcc Ni.}
\label{table:zdist}
\begin{ruledtabular}
\begin{tabular}{c|c|c}
                & {\tt SIESTA} & {\tt VASP} \\ \hline
$\Delta z_{12}$ & 2.011 & 2.008 \\
$\Delta z_{23}$ & 2.028 & 2.037 \\
$\Delta z_{34}$ & 2.021 & 2.039 \\
$\Delta z_{45}$ & 2.023 & 2.029 \\
$\Delta z_{56}$ & 2.011 & 2.007 \\
$a$             & 3.500 & 3.522   
\end{tabular}
\end{ruledtabular}
\end{center}
\end{table}

\begin{table}
\begin{center}
\caption{Atomically resolved magnetic moments for an isolated S-terminated Mn$_{12}$ and the AFM 
and FM configurations, in units of Bohr magneton ($\mu_B$), within the GGA, 
using {\tt VASP}. The numbers in the parentheses denote magnetic-moment differences between the 
isolated Mn$_{12}$ and the whole structure, where only the magnitudes are listed. 
The Mn ions are labeled in Fig.~\ref{fig:geo}(c). In the last row, the total magnetic moment 
of the Mn$_{12}$ is calculated from the sum of the atomically resolved magnetic moments.} 
\label{table:1}
\begin{ruledtabular}
\begin{tabular}{c|c|c|c}
Atomic species  & isolated S-terminated Mn$_{12}$   &   whole, AFM    &  whole, FM \\ \hline
Mn(1) or Mn(2)  & $-$2.63             &  2.62    (0.01)      & $-$2.60  (0.03)   \\
Mn(3) or Mn(4)  & $-$2.59             &  2.59    (0.00)      & $-$2.58  (0.01)  \\
Mn(5) or Mn(9)  &  3.53              & $-$3.54  (0.01)      &  3.54   (0.01)  \\
Mn(7) or Mn(11) &  3.52               & $-$3.54  (0.02)      &  3.54    (0.02)   \\
Mn(6) or Mn(10) &  3.50               & $-$3.56  (0.06)      &  3.57    (0.07)   \\
Mn(8) or Mn(12) &  3.56               & $-$3.57  (0.01)      &  3.57    (0.01)    \\
Mn (total)      & 17.78               & $-$18.00 (0.22)      & 18.05    (0.27)    \\
O closest to S  & $-$0.01,$-$0.01,$-$0.04,$-$0.04 & 0.00, 0.00, 0.00, 0.00 & 0.01, 0.01, 0.02, 0.02 \\
O  (total)      &  0.33               & $-$0.44               &  0.53              \\
C closest to S  & 0.01, 0.01         & 0.01, 0.01           & 0.00, 0.00        \\
C  (total)      &  0.21               & $-$0.18               &  0.19              \\
S(1) or S(2)    & $-$0.44             &  0.09               &  0.10              \\
Mn$_{12}$ (total)   & 17.44           & $-$18.41              & 18.95              
\end{tabular}
\end{ruledtabular}
\end{center}
\end{table}

\begin{table}
\begin{center}
\caption{The atomic sphere radii used for calculations of atomically resolved magnetic 
moments using {\tt VASP} (Table~\ref{table:1}).}
\label{table:WS}
\begin{ruledtabular}
\begin{tabular}{c|c}
Atomic species  &  Atomic radius (\AA)~ \\ \hline
Ni  &  1.29 \\
Mn  &  1.32 \\
S   &  1.16 \\
O   &  0.82 \\
C   &  0.86 \\
H   &  0.37  
\end{tabular}
\end{ruledtabular}
\end{center}
\end{table}

\begin{table}
\begin{center}
\caption{Energy difference ${\Delta E}$ (in eV) between the AFM and FM
configurations, within DFT and DFT+$U$ formalisms, using {\tt SIESTA} and {\tt VASP}.}
\label{table:2}
\begin{ruledtabular}
\begin{tabular}{c|c|c}
GGA,~~~{\tt SIESTA} & GGA,~~~{\tt VASP}  & GGA+$U$,~~~{\tt VASP} \\ \hline
$-$0.0395           &     $-$0.0516     &  $-$0.0463
\end{tabular}
\end{ruledtabular}
\end{center}
\end{table}

\begin{table}
\begin{center}
\caption{Total and induced magnetic moments (in $\mu_{\rm B}$) of the adsorbed Mn$_{12}$
for the AFM and FM configurations, within the GGA and GGA+U formalisms, using {\tt VASP}. 
The induced moments are obtained from comparison with the magnetic moment of an isolated 
S-terminated Mn$_{12}$.}
\label{table:Mn12}
\begin{ruledtabular}
\begin{tabular}{c|c|c|c|c}
     & Total (GGA) & Induced (GGA) & Total (GGA+U) & Induced (GGA+U) \\ \hline
AFM  &   $-$19.53  &  $-$1.53  &  $-$19.29 &  $-$1.29 \\
FM   &    20.31  &  2.31  &  20.23 &  2.23 
\end{tabular}
\end{ruledtabular}
\end{center}
\end{table}


\begin{thebibliography}{10}
\bibitem{ABINIT}
For construction of pseudopotentials, matching radii provided by the ABINIT team in the official 
SIESTA website were used.
\end{thebibliography}
\end{document}